\newcommand\etal{{\it et al.\/}\ }
\newcommand\ie{{\it i.e.\/}\rm\ }
\newcommand\eg{{\it e.g.\/}\rm\ }

\documentstyle[11pt,aaspp4]{article}
\lefthead{Scodeggio}
\righthead{The color-magnitude relation of early-type galaxies}

\begin{document}

\title{Internal color gradients and the color-magnitude relation
of early-type galaxies}

\author{Marco Scodeggio}
\affil{Istituto di Fisica Cosmica ``G. Occhialini'', via Bassini 15,
I-20133, Milano, Italy (marcos@ifctr.mi.cnr.it)}
%\email{marcos@ifctr.mi.cnr.it}

\begin{abstract}
The traditional use of fixed apertures in determining the well known
color-magnitude (CM) relation of early type galaxies, coupled with the
presence of radial color gradients within these systems, introduces a
bias in the CM relation itself. The effect of this bias is studied
here deriving a CM relation which is based on color measurements
carried out homogeneously within an aperture of radius equal to that
of the galaxy effective radius. A sample of 48 giant early-type
galaxies in the Coma cluster, with CCD observations in the $U$- and
$V$-band, is used for this derivation. It is found that internal
radial color gradients in early-type galaxies cannot be neglected when
discussing the colors of these systems, and that the CM relation
derived using color measurements within the effective radius is
significantly flatter than those based on fixed-aperture color
measurements. With the presently available data it is impossible to
determine whether the relation is completely flat, or whether a small
correlation is still present between galaxy color and luminosity.
\end{abstract}


\section{Introduction} 
\label{sec:intro}

It has long been known that the colors of early-type galaxies
correlate with their absolute magnitude (\eg \cite{B59}Baum 1959; 
\cite{dV61}de Vaucouleurs 1961; \cite{McC68}McClure \& van den Bergh 
1968; \cite{Las70}Lasker 1970). This color-magnitude (CM) relation was
demonstrated to be a universal property of early-type galaxies by
\cite{VS77}Visvanathan \& Sandage (1978), and has become one of the
most commonly used tools to constrain the epoch of formation and the
star formation history of these
systems. \cite{BLE92a}\cite{BLE92b}Bower, Lucey \& Ellis (1992a,b;
hereafter BLE92) were the first to study the CM relation using
photometric data based on CCD observations, which exhibited an even
smaller scatter around the mean relation than previous
determinations. More recently the CM relation has been shown to be
present in clusters up to $z \sim 1$ (\eg \cite{SED98}Stanford, Eisenhardt
\& Dickinson 1998; \cite{K98a}Kodama \etal 1998a; 
\cite{Gla98}Gladders \etal 1998), and to have similar properties in
distant and nearby clusters. 

These observations, when interpreted within the framework of a rapid
dissipative galactic collapse (\eg \cite{Lar74}Larson 1974), and with
the help of synthetic stellar population models (\eg
\cite{Buzz}Buzzoni 1989; \cite{BC93}Bruzual \& Charlot 1993; 
\cite{Wor94}Worthey 1994; \cite{Tan96}Tantalo \etal 1996; 
\cite{KA97}Kodama \& Arimoto 1997), yield an estimate for the
epoch of early-type galaxy formation of $z \sim 4.5$, with a fairly
limited star formation activity taking place afterwards
(\cite{Bow98}Bower, Kodama \& Terlevich 1998). The systematic change
of color as a function of luminosity in these models is interpreted as
the result of systematic changes of the mean metallicity as a function
of galaxy mass. These results currently provide what is probably the
strongest argument against the otherwise quite successful models of
hierarchical galaxy formation (but see \cite{KC98}Kauffmann \& Charlot
1998 for a different point of view).

The seemingly straightforward exercise of deriving a CM relation
observationally is complicated by the presence of systematic radial
color gradients within giant early-type galaxies.  It is now well
established that these objects are reddest in their cores, and become
bluer towards their peripheries (\eg \cite{dV61}de Vaucoleurs 1961; 
\cite{SV78}Sandage \& Visvanathan 1978; \cite{Fro78}Frogel \etal 
1978; \cite{Fra89}Franx, Illingworth \& Heckman 1989; 
\cite{Pel90a}Peletier \etal 1990a; \cite{Pel90b}Peletier, 
Valentijn \& Jameson 1990b), most likely as a result of radial
metallicity gradients within their stellar populations (see recent
analysis by \cite{Tam00}Tamura \etal 2000, and \cite{Sag00}Saglia 
\etal 2000).  Since the galaxy colors used in almost all studies of
the CM relation are measured within apertures of fixed radius,
different portions of a galaxy are used to derive these colors,
depending on the galaxy intrinsic size (and therefore on its
luminosity). As a result, a spurious correlation between galaxy
luminosity and color is produced, even in the absence of systematic
changes of the global stellar population as a function of galaxy
luminosity. This problem has been recognized for quite some time, and
a number of attempts have been made to take it into consideration.
For example \cite{SV78}Sandage \& Visvanathan (1978) corrected their
color measurements to a constant fraction of each galaxy isophotal
radius ($\theta / D(0) = 0.5$), but this correction does not solve the
problem, since isophotal radii include a variable fraction of galaxy
light, depending on galaxy luminosity. Other correction schemes have
been attempted on the modeling side, for synthetic stellar population
models (\eg \cite{K98a}Kodama \etal 1998a; \cite{KC98}Kauffmann \&
Charlot 1998). In this case mean radial color gradient values are used
to convert model-based total magnitudes into aperture ones, for a
direct comparison with the observations. However, given the large
scatter that is observed for the intensity of radial color gradients
(see for example \cite{Pel90a}Peletier \etal 1990a; or
section~\ref{sec:col_grad}), it is not appropriate to use a mean value
to characterize this property of the galaxy population. It would be
clearly preferable to derive observationally the CM relation using
variable-size apertures, measuring galaxy colors at a fixed fraction
of the galaxy light.

In this work a sample of 48 early-type galaxies in the Coma cluster,
with photometric observations in the $U$- and $V$-band and
measurements of the galaxy effective radius in the $V$- and $I$-band,
all available from the literature, is used to derive a CM relation
that is based on color measurements within the effective radius for
each galaxy in the sample. This newly obtained relation should be
directly comparable to the model predictions based on synthetic
stellar population models. The data-set used in this work is described
in section~\ref{sec:data}; the derivation of the traditional
``fixed-aperture'' and of the new ``fixed light fraction'' CM relation
are described in section~\ref{sec:cm_rel}; color gradients within
individual galaxies are discussed in section~\ref{sec:col_grad};
section~\ref{sec:conclusions} presents the discussion and the
conclusions of this work.

\section{The data}
\label{sec:data}

A sample of early-type galaxies in the Coma cluster was chosen for
this work, for a number of practical and astrophysical reasons. The
Coma cluster is among the richest nearby clusters, and has therefore a
large population of early-type galaxies. Due to its proximity,
accurate morphological information and a large set of photometric and
spectroscopic observations are available. Being a rich cluster, it
represents also a fair local counterpart to the distant clusters being
currently used to study the processes of formation and evolution of
early-type galaxies.  The Coma cluster is however too distant to allow
an expansion of the study of the CM relation to dwarf galaxies. This
exercise will be described in a future work (\cite{Sco01}Scodeggio 
\etal 2001), based on a sample of early-type galaxies in the Virgo cluster.

The dataset being used here consists of $U$- and $V$-band CCD
photometry observations of E and S0 galaxies in the Coma cluster,
taken from BLE92, supplemented by determinations of each galaxy
effective radius $r_e$ (the radius within which is contained half of
the galaxy light) taken from \cite{L91}Lucey \etal (1991) and from
\cite{SGH}Scodeggio, Giovanelli \& Haynes (1998; hereafter SGH98).
BLE92 obtained observations for 81 galaxies at the 2.5m Isaac Newton
Telescope (INT), in La Palma. For each galaxy these authors provide a
series of aperture magnitude measurements, carried out within
synthetic apertures with radius ranging from 4 to 60 arcsec.  The
typical accuracy of these measurements is 0.03 mag. Of these 81
objects, 56 were observed both in the $U$- and in the $V$-band, and
they are those used here. Total galaxy magnitudes in the $V$-band,
$V^T$, are also provided by BLE92, who re-calibrated the original data
from \cite{GP}Godwin \& Peach (1977).

Effective radius measurements for a subset of these galaxies were
presented by \cite{L91}Lucey \etal (1991), based in part on
independent $V$-band photometric observations also carried out at the
INT, and in part on BLE92 $V$-band data. The effective radius was
measured at the ``half light'' point of the growth curve of each
galaxy. Unfortunately, only 21 galaxies of the 56 with $U$- and
$V$-band photometry are in the \cite{L91}Lucey \etal sample. A much
larger fraction of these are in the $I$-band sample presented by
SGH98, based on CCD photometry observations carried out with the 0.9m
telescope of the Kitt Peak National Observatory.  These $r_e$
measurements are based on $r^{1/4}$ profile fits (see SGH98 for
details), that were extended from an inner radius twice as large as
that of the seeing disk to an outer radius chosen to match the radial
extent of the galaxy spheroidal component: for E galaxies this radius
is the one of the outermost reliable isophote, while for S0 galaxies
it is the radius where the disk component begins to dominate the
galaxy surface brightness.  Independent measurements for a number of
these galaxies were presented also by \cite{JFK95}J{\o}rgensen, Franx
\& Kj{\ae}rgaard (1995), and by \cite{Sag97}Saglia \etal (1997).  The
statistical uncertainty in the determination of $r_e$ is approximately
6\%, but significantly larger systematic uncertainties could be
produced by the particular fitting method used to derive $r_e$ and the
galaxy's effective surface brightness, leading to differences that, on
a given object, can be as large as 50\% (see, for example, the
discussion in SGH98). While a complete discussion on the reliability
of effective radius measurements is beyond the scope of the present
work, it is interesting to verify the possibility of using the larger
set of $I$-band $r_e$ measurements instead of the smaller set of
$V$-band ones, despite the significant wavelength difference. As
there is considerable overlap between the \cite{L91}Lucey \etal and
the SGH98 sample, it is possible to make a detailed comparison between
the two. The two sets of $r_e$ measurements appear to be in rather
good agreement, as shown in Figure ~\ref{fig:del_re}, where the
fractional difference $\Delta r_e / r_e$ is plotted as a function of
both $r_e$ and the galaxy total magnitude, for the 55 galaxies that
are common to the Lucey \etal and the SGH98 sample (this overlap
includes galaxies that are not in the BLE92 sample used here). The
average fractional difference is 0.092, with an {\it rms} scatter of
0.183, but a non-zero difference could be expected, as a consequence
of having used different photometric bands and different methodologies
to carry out the measurements. The important finding is that the
differences between $r_e$ measured in different bands do not depend on
galaxy luminosity or size, and therefore using $r_e$ as measured in
the $I$-band, as will be done hereafter (unless specifically stated)
does not introduce any luminosity or color related bias in the
following analysis.

For each galaxy, aperture magnitudes at $r_e$ were derived
independently in the two bands interpolating (or using a very small
extrapolation where necessary) the relative growth curve data. Given
the good sampling of each galaxy growth curves, and the fact that for
most galaxies the virtual aperture with radius $r_e$ is tightly
bracketed by a pair of aperture measurements, simple linear relations
between aperture magnitude and the logarithm of the aperture radius
were used for the interpolation/extrapolation. Only 8 of the 48
objects in the sample required an extrapolation of the measurements to
derive aperture magnitudes at $r_e$.  In all cases this was an
extrapolation to a radius smaller than the smallest aperture radius
used for the measurements (4 arcsec). This procedure was preferred to
the intuitively simpler one of just dividing by two the total flux
measured for a given galaxy (which would be equivalent to defining
$m(r_e) = m_T + 0.753$), because the latter always requires a
relatively large extrapolations to be carried out on the measurements
to derive total fluxes. This extrapolation process would undoubtedly
become the largest source of color measurement uncertainty, while the
procedure adopted here only adds to the original photometric
uncertainty the uncertainty on the determination of $r_e$, which is
significantly smaller (see section~\ref{sec:cm_re}).  Galaxy colors
within $r_e$ were obtained taking the difference between the relative
aperture magnitudes. As a consistency check colors were also obtained
by interpolating/extrapolating a linear relation between color and the
logarithm of the aperture radius. Differences between the two
techniques show an {\it rms} scatter of 0.018 mag, when only
interpolation between observations is required to derive the color
within $r_e$. Only colors derived from the difference of aperture
magnitudes are used in the following analysis.

Among the 55 objects with $r_e$ measurements both in $V$- and
$I$-band, 20 are also in the BLE92 sample, and for them it was
possible to obtain a color measurement within apertures of radius
$r_e$ as determined in the two bands. The differences between these
two measurements are shown in Figure~\ref{fig:del_color}, as a
function of galaxy magnitude, effective radius, and color. As could be
expected from the above discussion on the differences between
effective radii measured in the two bands, no significant difference
is present between the two sets of color measurements, except for a
small offset introduced by the somewhat larger values of $r_e$ as
determined in the $I$-band. The average difference between the $U-V$
colors is -0.018 mag, with an {\it rms} scatter of 0.032 mag. The
measured scatter is in good agreement with the expected uncertainty in
the color measurements introduced by measurement errors on $r_e$ (see
also Section~\ref{sec:cm_re}). The fact that the color difference is
independent from galaxy luminosity and color means that no bias is
introduced in the $[(U-V),~V]$ CM relation by the choice of measuring
colors within an aperture of radius the $I$-band effective radius.

The sample used in the following analysis is therefore composed of 48
galaxies with magnitude measurements both in $U$- and $V$-band, and a
measurement of $r_e$ in the $I$-band. Morphological types for all
objects are available from \cite{D80}Dressler (1980): 23 of them are
classified as Ellipticals, and 25 as S0. The sample is limited to
galaxies brighter than $V\simeq 15.8$, but does not obey any strict
completeness criterion. Assuming a distance modulus for Coma of 34.75,
the limiting magnitude corresponds to $M_V\simeq -19.0$, or
equivalently $M_B \simeq -18.0$.

\section{The Color-Magnitude relation}
\label{sec:cm_rel}

\subsection{The standard ``fixed-aperture'' relation}
\label{sec:cm_fixed}

The standard ``fixed-aperture'' CM relation was derived using two
slightly different approaches. First, the relation was obtained in the
form that is most often used in the analysis of high-redshift clusters,
\ie using both color and magnitude measurements within a given fixed
aperture (\eg \cite{Bow98}Bower \etal 1998; \cite{K98b}Kodama, Bower
\& Bell 1998b; \cite{Gla98}Gladders \etal 1998). In the second
approach, which was introduced to facilitate the comparison with the
``fixed light fraction'' relation discussed in
Section~\ref{sec:cm_re}, colors measured in fixed apertures were
correlated with magnitudes measured within the aperture of radius
$r_e$ and with total magnitudes.  This approach is very similar to the
one used by BLE92 and \cite{SED98}Stanford \etal (1998).

The two procedures produce very similar results, except for the fact
that using a fixed aperture both for the color and the magnitude
measurement introduces the bias related to sampling different portions
of a galaxy depending on its luminosity (and therefore also on its
color) in both quantities, and this makes the observed CM relation
steeper. This is illustrated in Figure~\ref{fig:cm_fixed_ap}, where
the $[(U-V),~V]$ CM relation obtained when measuring the colors within
an aperture of 10 arcsec, and the magnitudes within the same 10 arcsec
aperture, within $r_e$, or when using total magnitudes, are compared.
The slope obtained when using total magnitudes is $-0.074 \pm 0.008$,
in good agreement with the one previously reported by BLE92 ($-0.082
\pm 0.008$).

One should also notice that the slope of the ``fixed-aperture'' CM
relation is marginally related to the size of the fixed aperture used
for the color measurements, becoming slightly flatter for larger
aperture radii. This flattening, although not statistically
significant in the present sample, is in good agreement with a similar
result reported by \cite{Oka99}Okamura \etal (1999), based on $B-R$ color
measurements also in the Coma cluster.

Table~\ref{tab:cm_summary} presents a summary of the linear least
square fits to the CM relations discussed in this work. All fits are
direct fits, where the fitting procedure minimizes the vertical
distances between data-points and the best-fit line, taking into
account only error measurements in the independent variable, the color
in this case. This fitting procedure was adopted because for the total
magnitude measurements individual uncertainty estimates are not
available, and also because the main goal of this paper is to
illustrate the relative changes in the CM relation when different
methods are used to derive it, more than to derive the ultimate value
of the relation parameters. It is certainly possible that such a very
simplistic approach to the fitting of the CM relation might bias the
result towards flatter slopes, especially when measurements errors are
large. This might be happening within the present dataset, as shown by
the comparison between the correlation involving magnitudes measured
within a 10 arcsec aperture (first line in the Table; accurate
measurements) and the correlation involving total magnitudes or
magnitudes within $r_e$ (second and third line in the Table; less
accurate measurements).  Uncertainties in the value of the best-fit
slope and zero point were obtained with the statistical jackknife
method. With a sample of $N$ data-points, the fit is repeated $N$
times using $N-1$ points, each time excluding a different point, to
derive an estimate of the parent population from which the parameter
under examination is derived. With the present dataset, formal
uncertainties on the best-fit slopes are approximately one order of
magnitude smaller than those based on the statistical
jackknife. However, one must keep in mind that the two given
uncertainties are highly correlated, and cannot be used independently
to asses the global uncertainty affecting the fit. An approximate
estimate of the zero point uncertainty component that is statistically
independent from the slope uncertainty is given by the accuracy with
which one can measure the average value of the residuals from the
best-fit line, which is typically of 0.02 mag. The scatter in the CM
relation reported in the last column of the table is the {\it rms}
scatter of the differences between the measured color and the one
predicted by the fit.

\subsection{The relation at a fixed galaxy light fraction}
\label{sec:cm_re}

As discussed in Section~\ref{sec:intro}, if there are radial color
gradients within galaxies, the use of fixed apertures for color
measurements introduces a bias in the CM relation, as more luminous
galaxies have larger $r_e$ (see, for example, \cite{F64}Fish 1964; 
\cite{Guz93}Guzm{\'a}n, Lucey \& Bower 1993; \cite{Pah98}Pahre,
Djorgovski \& de Carvalho 1998), and therefore larger overall extent.
Since radial color and line strength gradients are commonly present
within giant early-type galaxies (see, for example, \cite{dV61}de 
Vaucoleurs 1961; \cite{SV78}Sandage \& Visvanathan 1978; 
\cite{Fro78}Frogel \etal 1978; \cite{Fra89}Franx \etal 1989; 
\cite{Pel90a}\cite{Pel90b}Peletier \etal 1990a,b on the color 
gradients, and \cite{CH88}Couture \& Hardy 1988; \cite{Gor90}Gorgas,
Efstathiou \& Arag{\'o}n-Salamanca 1990; \cite{Dav93}Davies, Sadler \&
Peletier 1993; \cite{CD94}Carollo \& Danziger 1994 on the line
strength ones) one can expect the ``fixed-aperture'' CM relation
discussed in the previous section to provide a biased view of the
color properties of early-type galaxies.

To quantify the relevance of this effect, a ``fixed light fraction''
CM relation was derived, using the colors measured within each galaxy
effective radius. With this procedure, always a fixed fraction of the
galaxy light contributes to the color measurements, allowing a more
meaningful comparison between galaxies of different luminosities.
Contrary to what appeared to be the implicit assumption in most
previous works on the CM relation, the bias introduced by the use of
fixed apertures appears to be significant. Figure~\ref{fig:cm_comp}
shows a comparison of the CM relations obtained when using color
measurements within a fixed aperture of 10 arcsec, and within
$r_e$. It is clear that any correlation between galaxy color and
luminosity largely disappears when colors are measured within each
galaxy effective radius. The slope of the best fitting linear relation
between color and magnitude goes from $-0.074 \pm 0.008$ (as measured
for the fixed aperture relation) to $-0.016 \pm 0.018$, which is
statistically compatible with a completely flat relation.  A similarly
shallow slope for the CM relation derived measuring galaxy colors
homogeneously at the effective radius was previously reported by
\cite{PS96}Prugniel \& Simien (1996) and by \cite{PvD98}van Dokkum
\etal (1998), although those authors do not comment specifically on
the significance of this point.

Some concern about the reality of this result might be raised by the
fact that color measurements within $r_e$ are affected by larger
uncertainties than those within a fixed radius aperture, because of
the uncertainty in the determination of $r_e$ itself. As discussed in
Section~\ref{sec:data}, the statistical uncertainty in the
determination of $r_e$ for the present sample is approximately 6\%
(SGH98), but much larger systematic uncertainties could be present,
introduced by the specific fitting procedure used to derive the value
of $r_e$ from a galaxy surface brightness profile (see, for example,
Figures 4 to 6 in SGH98). Therefore a total uncertainty of
approximately 20--30\% might be a better estimate of the error budget
involved in the determination of $r_e$.  However, the impact on the
present color measurements of a conservative 20\% $r_e$ uncertainty
estimate (the scatter between $V$- and $I$-band measurements discussed
in Section~\ref{sec:data} would lead to an estimate of approximately
10--12\%) is only $\pm 0.025$ mag, on average. Since this uncertainty
is always smaller than, or comparable to the intrinsic uncertainty of
the presently available photometric measurements, it does not have a
significant impact on the results presented here. The 0.03 mag
uncertainty quoted by BLE92 for their magnitude measurements
translates into a 0.042 mag uncertainty in the fixed-aperture color
determinations, and the addition of the contribution from $r_e$
determination uncertainties brings the total fixed galaxy fraction
color uncertainty to approximately 0.05 mag. These uncertainties are
reported in Figure~\ref{fig:cm_comp}, together with a somewhat
arbitrary estimate of 0.15 mag for the uncertainty in the total
magnitude measurements.

\section{Radial color gradients}
\label{sec:col_grad}

The significant changes in the CM relation properties described in the
previous section point towards the relevance of internal color and
stellar population gradients within early-type galaxies for a detailed
understanding of their properties and evolutionary histories.  The
average color gradient $d(U-V)/d(\log r)$ measured in the sample used
here is of $-0.15$ mag per dex in radius, with an {\it rms} scatter
around this mean value of 0.15 mag per dex in radius.  This is in very
good agreement with the findings of \cite{Pel90a}Peletier \etal
(1990a), who report an average gradient of $-0.16 \pm 0.11$ mag per
dex in radius. Note that the use, in both cases, of a simple power-law
method to measure the color gradients makes the comparison meaningful,
although different methods are used to measure the color points.

In agreement with previous studies, no significant correlation is
found between the strength of the color gradient and the galaxy
luminosity (see Figure~\ref{fig:color_grad}). There is however a
marginal indication that the scatter in gradient strength might be
smaller for the brighter objects in our sample. This result is
somewhat in contrast with the claim presented by \cite{Pel93}Peletier
(1993) that early-type galaxies fainter than M$_B = -19.5$ have
significantly smaller color gradients than those observed in more
luminous galaxies.  Unfortunately the number of bright galaxies in the
present sample is too small to allow a clear determination of this
effect.

The rather large scatter that is observed globally for the values of
the color gradient is responsible for the larger scatter measured in
the fixed galaxy fraction version of the CM relation with respect to
the fixed aperture version. This fact points towards a very high
photometric accuracy requirement for future measurements aimed at
measuring with good accuracy the slope of the CM relation. Also, it
becomes quite clear that using an average gradient value to correct
color measurements derived from either observations or theoretical
models is not appropriate. In fact the scatter in gradient values
observed among real galaxies is comparable in magnitude with the mean
value itself, instead of representing a small additional uncertainty
superposed to the main effect one is considering. Therefore when using
a mean gradient value for the color corrections one would neglect the
effects of the relatively strong gradients that are present in a
significant fraction of the galaxy population.

\section{Discussion and Conclusions}
\label{sec:conclusions}

The results presented above show that internal radial color gradients
in early-type galaxies cannot be neglected when discussing the colors
of these systems. The strength of the gradients measured here is in
agreement with other measurements previously reported in the
literature, although differences in samples and in the photometric
bands used by different authors make a detailed comparison
difficult. It is confirmed that for giant early-type galaxies the
strength of the color gradient is not correlated with the galaxy
luminosity. Also, a fairly large scatter in this strength is observed
at all luminosities, although there is an indication in the present
dataset that this scatter might be smaller for the brightest objects.

One point of concern with the present dataset is the rather
heterogeneous nature of the observations and data reduction procedures
adopted to derive magnitudes and effective radii. However one should
note that a substantial flattening of the CM relation when colors are
measured consistently at the galaxies effective radius, is obtained
also when considering $I$- and $H$-band data, with effective radii
measured homogeneously within the same data set (\cite{Sco01}Scodeggio 
\etal 2001). Very shallow ``fixed light fraction'' CM relation slopes
have been published also by \cite{PS96}Prugniel \& Simien (1996) and 
\cite{PvD98}van Dokkum \etal (1998).

While it is found that colors measured within the effective radius
depend less strongly on the galaxy luminosity than previously
believed, it is not possible to accurately determine whether a small
correlation is still present between the two quantities, or whether
giant E and S0 galaxies all have the same color, within a small
scatter. Because only small differences in color are involved, a
larger, more homogeneous dataset, comprising data of higher
photometric accuracy than those used here, will be required to settle
this point. In any case the present result implies that some
modification of the models of galaxy formation currently discussed in
the literature is necessary. The shallower slope (if not the complete
flatness) of the CM relation, and the consequential somewhat larger
scatter allowed in the colors of early-type galaxies with respect to
that measured in the fixed-aperture relation, should significantly
relax some of the tight constraints so far taken into account in
discussing the allowed star formation histories of these systems and
the processes that led to their formation. 

The present results apply only to giant E and S0 galaxies, since with
this dataset one can sample only the bright end of the early-type
galaxies luminosity function (spanning an interval of approximately
five magnitudes). The picture appears to become more complex when one
considers also dwarf systems. We are currently extending our analysis
to the Virgo cluster, to build a sample that includes also dE and dS0
galaxies (Scodeggio \etal 2000). Our preliminary finding is that a
global blueing trend with decreasing luminosity is present among
early-type galaxies. The trend is significant for the global sample of
Virgo cluster galaxies, spanning an interval of more than ten
magnitudes in luminosity, but becomes vanishingly small at the bright
end of the distribution, in agreement with the results presented
here. This would suggest the possibility of a CM relation which is
divided into two different regimes, one for the giant and one for the
dwarf galaxies.

It would be quite natural to extend the conclusions reached here to
other well known scaling relations for early-type galaxies, like the
metallicity-luminosity relation, exemplified primarily by the
Mg$_2$-$\sigma$ relation (\eg \cite{BBF}Bender, Burstein \& Faber
1993). In this case once again the measurements are made using a fixed
aperture. The typical size of the aperture is comparable with the
galaxy $r_e$ for the small, low luminosity and low $\sigma$ galaxies,
while it is one order of magnitude smaller than $r_e$ in the case of
the large, luminous, high $\sigma$ ones. Since there are radial
gradients in the strength of the Mg$_2$ index, the approach may
introduce a significant bias in the inferred mass-metallicity
relation. However a recent analysis of a relatively large set of
measurements of such gradients has led \cite{KA99}Kobayashi \& Arimoto
(1999) to conclude that a metallicity-mass relation is present among
elliptical galaxies even when mean metallicities are considered.

The conclusion is that the interpretation of the CM relation of bright
early-type galaxies, as well as that of similar relations with galaxy
luminosity that might suffer from the same ``fixed-aperture plus
internal gradient'' bias, might need to be revised. This could lead to a
significant change in the overall observational picture of early-type
galaxies.  In particular it could no longer be required to have a very
strong correlation between galaxy luminosity (which is approximately
equivalent to mass) and mean metallicity, allowing for a more
significant contribution from random events (like mergers, or a number
of star formation episodes besides the original burst) in the
formation and evolutionary processes that shape these objects.

\acknowledgments
The most sincere thanks to Alessandro Boselli, Peppo Gavazzi, Riccardo
Giovanelli, and Martha Haynes for their contributions to this project,
and to Stephane Charlot, Mark Dickinson, Alvio Renzini, Piero Rosati,
and Roberto Saglia for useful discussions on the subject.

\newpage

\begin{deluxetable}{cccc}
\tablewidth{0pt}
\tablenum{1}
\tablecaption{Derived $[(U-V),~V]$ color-magnitude relations 
\label{tab:cm_summary}}
\tablehead{
\colhead{Apertures\tablenotemark{a}} & \colhead{slope} & \colhead{zero point}
& \colhead{scatter (mag)}
}
\startdata
$10\arcsec$,~$10\arcsec$ & $-0.119 \pm 0.026$ & $3.28 \pm 0.40$ & 0.080\nl
$10\arcsec$,~$r_e$       & $-0.072 \pm 0.012$ & $2.50 \pm 0.21$ & 0.086\nl
$10\arcsec$,~tot.        & $-0.074 \pm 0.008$ & $2.52 \pm 0.11$ & 0.059\nl
$r_e$,~tot.              & $-0.016 \pm 0.018$ & $1.73 \pm 0.28$ & 0.136\nl
\enddata
\tablenotetext{a}{the format for this column is (aperture used for the
color measurement),(aperture used for the magnitude measurement).}
\end{deluxetable}
\newpage

\pagestyle{empty}

%\newpage

%\voffset=2truecm
%\hoffset=-1.2truecm
\
\plotone{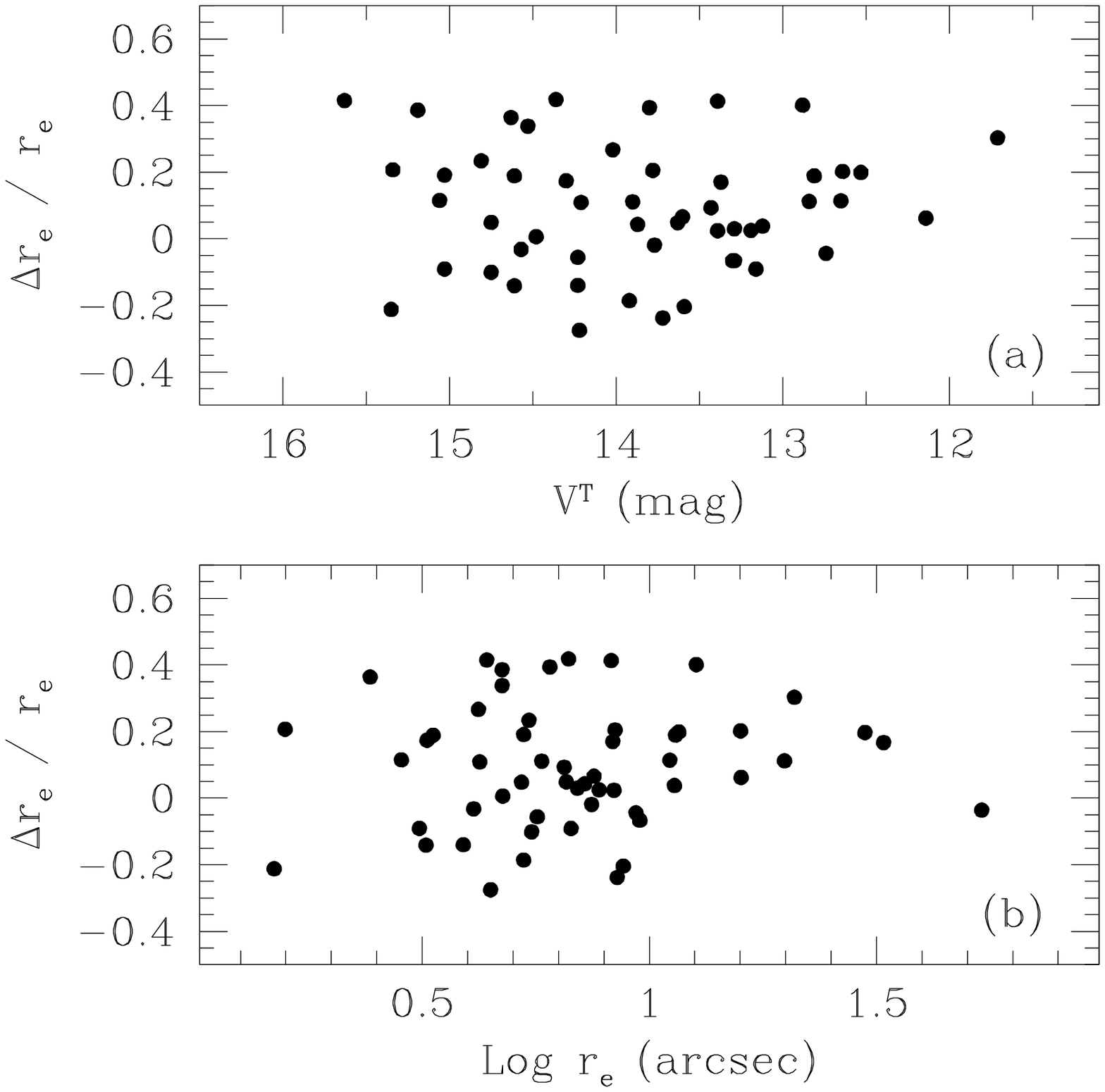}
\figcaption[del_re.ps]{Differences in the values of effective radius
as measured in the $V$- (Lucey \etal 1991) and $I$-band (Scodeggio,
Giovanelli \& Haynes 1998). In the two panels the ratio of the
difference between the $r_e$ measurements in the $I$- and $V$-band
with respect to their mean is plotted as a function of the
galaxy total magniude (panel a) or the logarithm of the $I$-band
measurement (panel b). \label{fig:del_re}}

\plotone{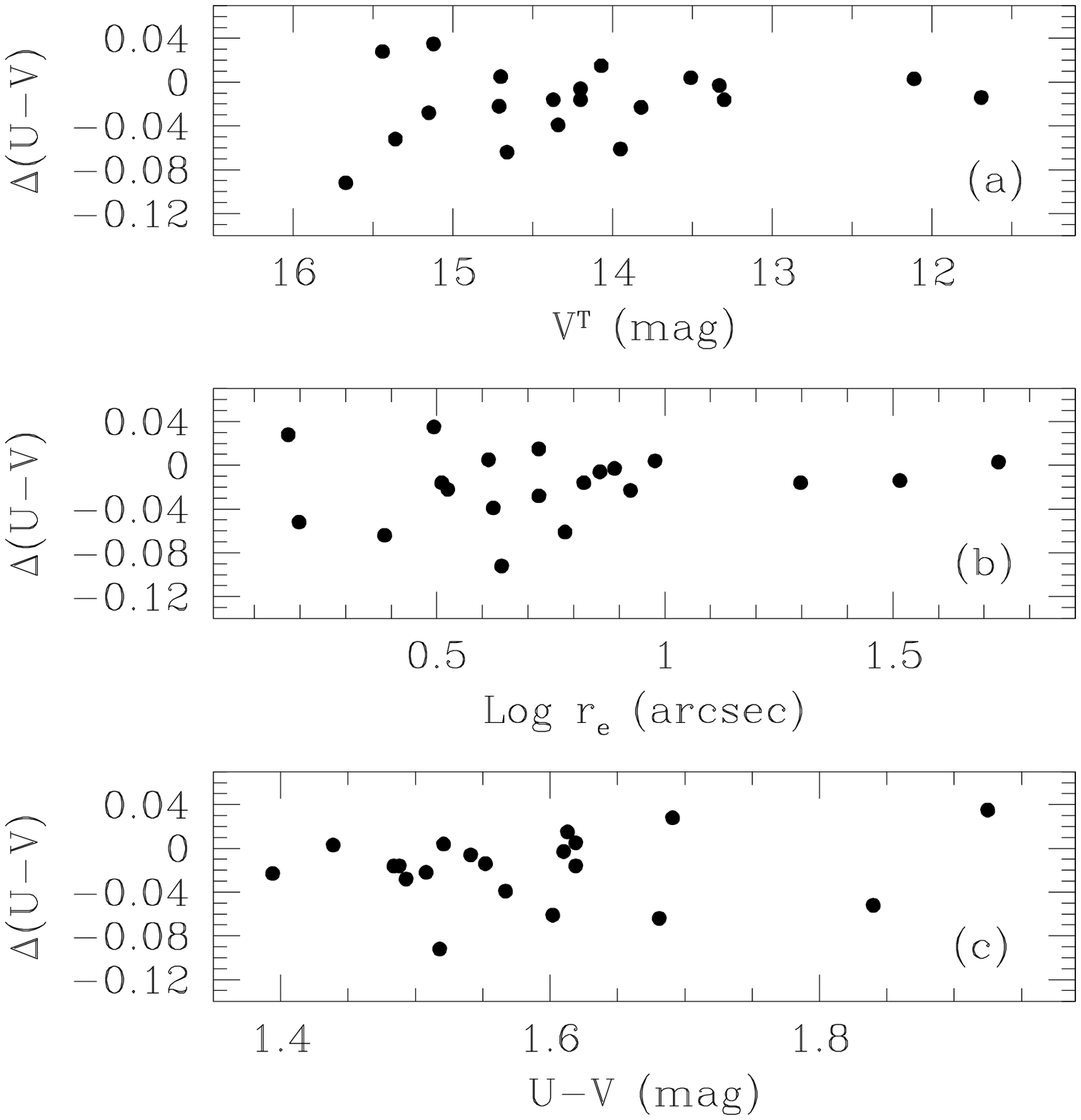}
\figcaption[del_color.ps]{Differences in the values of the $U-V$
color, as measured within an aperture of radius the $V$- and $I$-band
effective radius. In the three panels the difference between the color
measured within the $I$-band and the $V$-band $r_e$ is plotted as a
function of the galaxy total magnitude (panel a), the logarithm of the
$I$-band effective radius (panel b), and the $U-V$ color measured
within that radius (panel c). \label{fig:del_color}}

\plotone{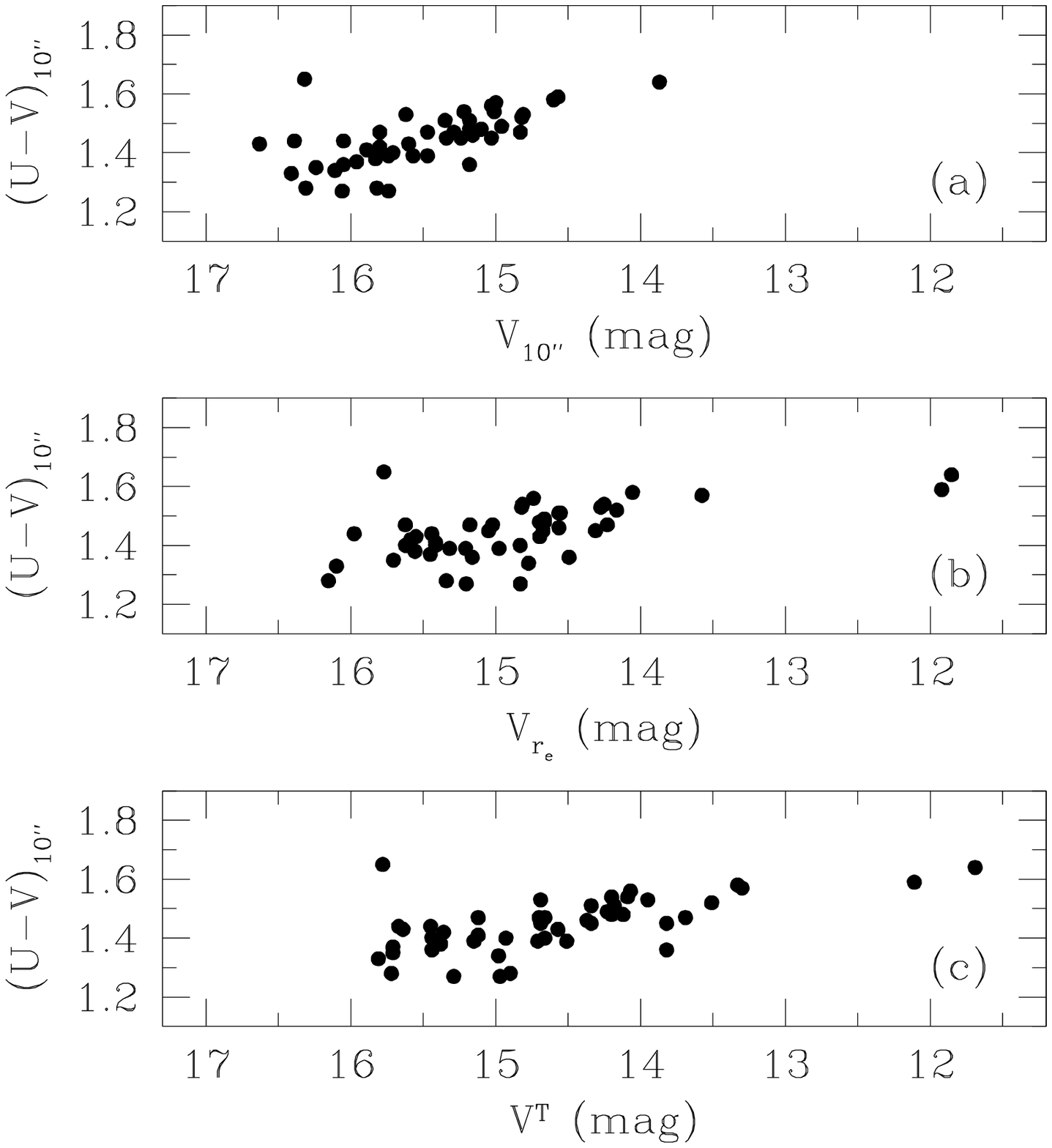}
\figcaption[cm_fixed.ps]{``Fixed-aperture'' color-magnitude
relations. The $[(U-V),V]$ relation is plotted in three different
formats: in panel (a) both color and magnitude are measured within a
10 arcsec aperture; in panels (b) and (c) the color is the same as in
panel (a), but the magnitudes are the one measured within an aperture
of radius the galaxy effective radius (panel b) and the total
magnitude (panel c). \label{fig:cm_fixed_ap}}

\plotone{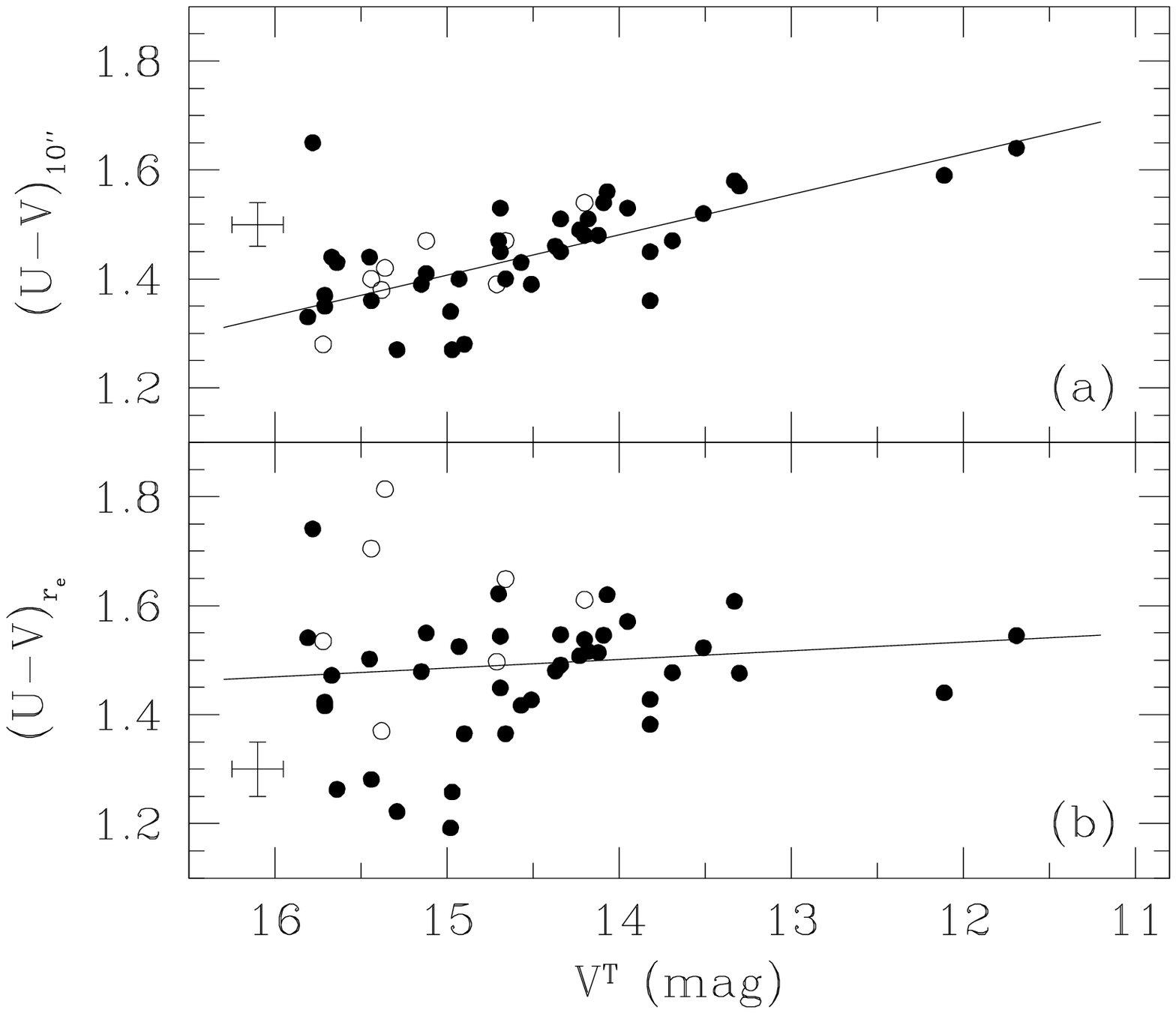}
\figcaption[cm_comp.ps]{Comparison between ``fixed-aperture'' and ``fixed
light fraction'' color-magnitude relations. Panel (a) refers to the
color measured within a fixed 10 arcsec aperture, while the panel (b)
refers to the color measured within the galaxy effective radius. In
both cases the magnitude is the galaxy total magnitude. Empty symbols
identify those galaxies that required an extrapolation of their growth
curve to derive the color measurement at the effective radius. The
solid lines represent the best fit to the relation discussed in the
text. The ``crosses'' identify typical uncertainties in the plotted
colors and magnitudes. \label{fig:cm_comp}}

\plotone{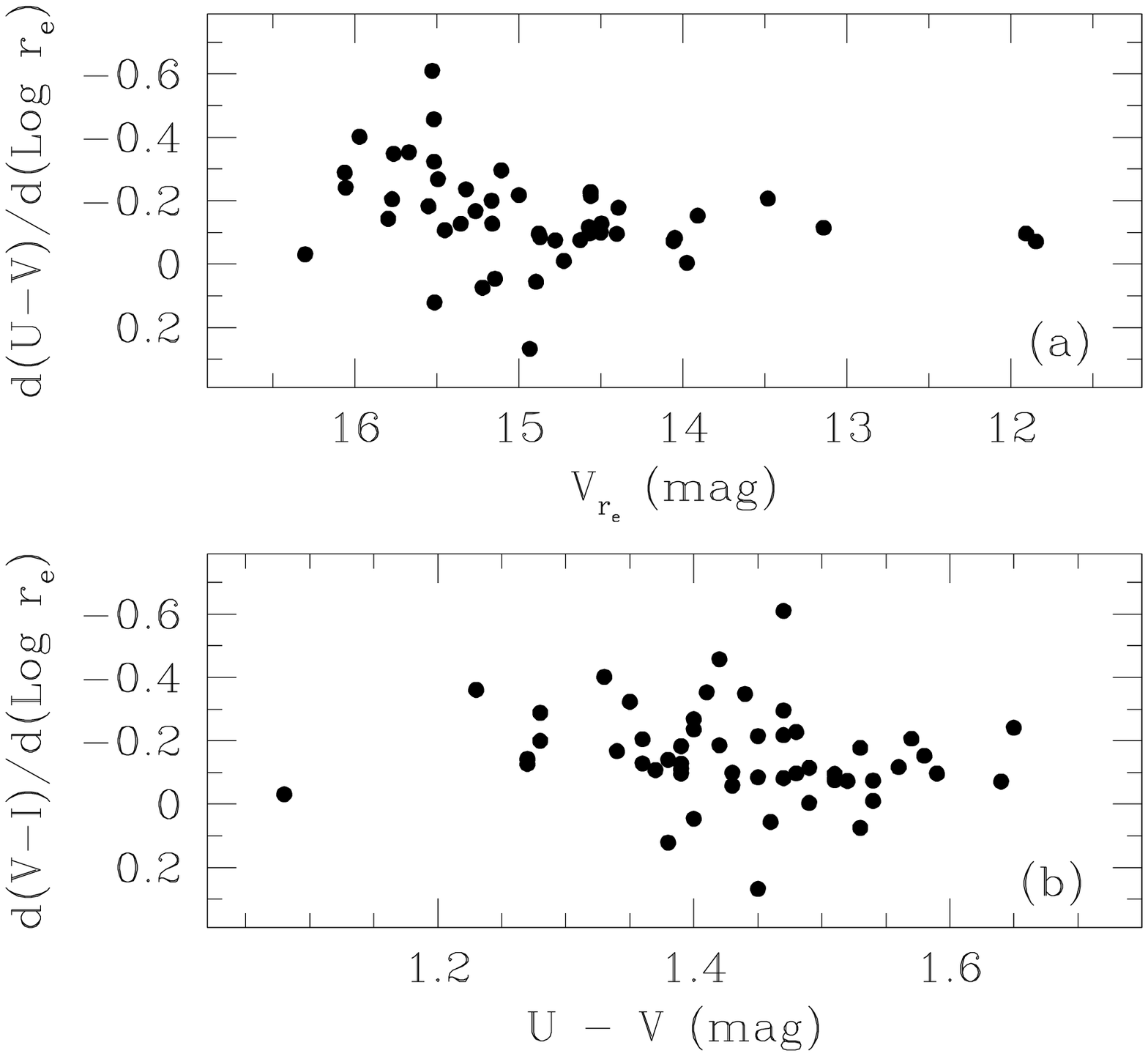}
\figcaption[col_grad.ps]{The strength of the radial internal color
gradient $d(U-V)/d \log(r)$, as a function of galaxy luminosity
(represented here by the integrated magnitude within the galaxy
effective radius; panel a), and galaxy color (panel b).
\label{fig:color_grad}} 


\begin{thebibliography}{}

\bibitem[]{B59} Baum, W.A., 1959, PASP, 71, 106
\bibitem[]{BBF} Bender, R., Burstein, D., Faber, S.M., 1993, ApJ, 411, 153 
\bibitem[]{BLE92a} Bower, R.G., Lucey, J.R., Ellis, R.S., 1992a, MNRAS, 
	254, 589 (BLE92)
\bibitem[]{BLE92b} Bower, R.G., Lucey, J.R., Ellis, R.S., 1992b, MNRAS, 
	254, 601 (BLE92)
\bibitem[]{Bow98} Bower, R.G., Kodama, T., Terlevich, A., 1998, MNRAS,
	299, 1193 
\bibitem[]{BC93} Bruzual, A.G., Charlot, S., 1993, ApJ, 405, 538 
\bibitem[]{Buzz} Buzzoni, A., 1989, ApJS, 71, 817
\bibitem[]{CD94} Carollo, C.M., Danziger, I.J., 1994, MNRAS, 270, 523
\bibitem[]{CH88} Couture, J., Hardy, E., 1988, AJ, 96, 867
\bibitem[]{Dav93} Davies, R.L., Sadler, E.M., Peletier, R.F., 1993,
	MNRAS, 262, 650 
\bibitem[]{dV61} de Vaucouleurs, G., 1961, ApJS, 5, 233
\bibitem[]{D80} Dressler, A., 1980, ApJS, 42, 565
\bibitem[]{F64} Fish, R.A., 1964, ApJ, 139, 284
\bibitem[]{Fra89} Franx, M., Illingworth, G.D., Heckman, T., 1989, AJ, 98, 538
\bibitem[]{Fro78} Frogel, J.A., Persson, S.E., Aaronson, M., Matthews, K., 
	1978, ApJ, 220, 75
\bibitem[]{Gla98} Gladders, M.D., Lopez-Cruz, O., Yee, H.K.C., Kodama, T.,
	1998, ApJ, 501, 571
\bibitem[]{GP} Godwin, J.G., \& Peach, J.V., 1977, MNRAS, 181, 323
\bibitem[]{Gor90} Gorgas, J., Efstathiou, G., Arag{\'o}n-Salamanca, A., 1990,
	MNRAS, 245, 217
\bibitem[]{Guz93} Guzm{\'a}n, R., Lucey, J.R., Bower, R.G., 1993,
	MNRAS, 265, 731 
\bibitem[]{KC98} Kauffmann, G., Charlot, S., 1998, MNRAS, 294, 705
\bibitem[]{KA99} Kobayashi, C., Arimoto, N., 1999, ApJ, 527, 573
\bibitem[]{KA97} Kodama, T., Arimoto, N., 1997, A\&A, 320, 41
\bibitem[]{K98a} Kodama, T., Arimoto, N., Barger, A.J., 
	Arag{\'o}n-Salamanca, A., 1998a, A\&A, 334, 99
\bibitem[]{K98b} Kodama, T., Bower, R.G., Bell, E.F., 1998b, astro-ph/9810138
\bibitem[]{JFK95} J{\o}rgensen, I., Franx, M. \& Kj{\ae}rgaard, P. 1995, 
	MNRAS, 273, 1097
\bibitem[]{Lar74} Larson, R.B., 1974, MNRAS, 166, 585
\bibitem[]{Las70} Lasker, B.M., 1970, AJ, 75, 21
\bibitem[]{L91} Lucey, J.R., Guzm{\'a}n, R., Carter, D., Terlevich, R.J.,
	1991, MNRAS, 253, 584
\bibitem[]{McC68} McClure, R.D., van den Bergh, S., 1968, AJ, 73, 1008
\bibitem[]{Oka99} Okamura, S., et al., 1999, in ``Cosmological Parameters
	and the Evolution of the Universe'', ed. Sato, K., (Kluwer:
	Dordrecht), p. 160 
\bibitem[]{Pah98} Pahre, M.A., Djorgovski, S.G., de Carvalho, R.R., 1998, 
	AJ, 116, 1591
\bibitem[]{Pel93} Peletier, R.F., 1993, in ``Structure, Dynamics, and Chemical
	Evolution of Elliptical Galaxies'', ed. Danziger, I.J., Zeilinger, 
	W.W., Kjaer, K., (ESO: Garching) p. 409
\bibitem[]{Pel90a} Peletier, R.F., Davies, R.L., Illingworth, G.D., Davis,
	L.E., Cawson, M., 1990a, AJ, 100, 1091
\bibitem[]{Pel90b} Peletier, R.F., Valentijn, E.A., Jameson, R.F., 1990b,
	A\&A, 233, 62
\bibitem[]{PS96} Prugniel, P., Simien, F., 1996, A\&A, 309, 759
\bibitem[]{Sag97} Saglia, R.P., et al., 1997, MNRAS, 292, 499
\bibitem[]{Sag00} Saglia, R.P., Maraston, C., Greggio, L., Bender, R.,
	Ziegler, B., 2000, A\&A, 360, 911
\bibitem[]{SV78} Sandage, A., Visvanathan, N., 1978, ApJ, 223, 707
\bibitem[]{SGH} Scodeggio, M., Giovanelli, R., Haynes, M.P., 1998, AJ,
	116, 2728 (SGH98)
\bibitem[]{Sco01} Scodeggio, M., et al., 2001, in preparation
\bibitem[]{SED98} Stanford, S.A., Eisenhardt, P.R., Dickinson, M., 1998, 
        ApJ, 492, 461
\bibitem[]{Tam00} Tamura, N., Kobayashi, C., Arimoto, N., Kodama, T., Ohta,
	K., 2000, AJ, 119, 2134
\bibitem[]{Tan96} Tantalo, R., Chiosi, C., Bressan, A., Fagotto, F., 1996,
	A\&A, 311, 361
\bibitem[]{PvD98} van Dokkum, P.G., Franx, M., Kelson, D.D.,
	Illingworth, G.D., Fisher, D., Fabricant, D., 1998, ApJ, 500, 714
\bibitem[]{VS77} Visvanathan, N., Sandage, A., 1977, ApJ, 216, 214
\bibitem[]{Wor94} Worthey, G., 1994, ApJS, 95, 107

\end{thebibliography}
\end{document}